\documentclass[prl,twocolumn,showpacs,preprintnumbers,amsmath,amssymb]{revtex4-1} 
\usepackage{dcolumn}% Align table columns on decimal point
\usepackage{bm}% bold math

\usepackage{amsmath}
\usepackage{amssymb}
\usepackage{amsfonts} 
\usepackage{wasysym}  
\usepackage{graphics}  
\usepackage{psfrag} 
\usepackage{graphicx} 
\usepackage{epsfig}    
\usepackage{rotating}  
 \usepackage[latin1]{inputenc}
 \usepackage{color}
 \usepackage{rotating}
 \usepackage{multirow}
 \usepackage{footnote}

\usepackage{color}
\usepackage{epsfig}

\usepackage{tikz}
\usetikzlibrary{arrows,shapes,positioning}
\usetikzlibrary{decorations.pathmorphing} % noisy shapes
\usetikzlibrary{fit}			% fitting shapes to coordinates
\usetikzlibrary{backgrounds}

\newcommand{\mean}[1]{\left\langle #1 \right\rangle}

\newcommand{\beq}{\begin{equation}}
\newcommand{\eeq}{\end{equation}}
\newcommand{\beqn}{\begin{eqnarray}}
\newcommand{\eeqn}{\end{eqnarray}}
\renewcommand{\vec}{\mathbf}
\newcommand{\Ker}[1]{\text{Ker}\left( #1\right) }

\newcommand{\Pf}[1]{\text{Pf}\left( #1\right) }

\definecolor{darkgreen}{rgb}{0,0.5,0} 
\definecolor{violet}{rgb}{0.5,0,0.5}
\definecolor{orange}{rgb}{0.8,0.5,0.2}
\definecolor{grey}{rgb}{0.3,0.3,0.3}

\begin{document}

%\preprint{LMU-ASC 63/12}

\title{Coexistence and Survival in Conservative Lotka-Volterra Networks} 

\author{Johannes Knebel$^{1}$}
\author{Torben Kr\"uger$^{2}$}
\author{Markus F. Weber$^{1}$}
\author{Erwin Frey$^{1}$}
\email{frey@lmu.de}

\affiliation{
$^{1}$Arnold Sommerfeld Center for Theoretical Physics and Center for NanoScience, Department of Physics, Ludwig-Maximilians-Universit\"at M\"unchen, Theresienstrasse 37, 80333 M\"unchen, Germany\\
$^{2}$Department of Mathematics, Ludwig-Maximilians-Universit\"at M\"unchen, Theresienstrasse 38, 80333 M\"unchen, Germany}

\begin{abstract}
Analyzing coexistence and survival scenarios of Lotka-Volterra (LV) networks in which the total biomass is conserved is of vital importance for the characterization of long-term dynamics of ecological communities.
Here, we introduce a classification scheme for coexistence scenarios in these conservative LV models and quantify the extinction process by employing the Pfaffian of the network's interaction matrix.
We illustrate our findings on global stability properties for general systems of four and five species and find a generalized scaling law for the extinction time.
\end{abstract}


\maketitle

%%%%%%%%%%%%%%%%
%General introduction
%%%%%%%%%%%%%%%%

Understanding the stability of ecological networks is of pivotal importance in theoretical biology~\cite{May, Montoya2006}. Coexistence and extinction of species depend on many factors such as inter- and intra-species interactions~\cite{Szabo2007b, Mathiesen2011}, population size~\cite{Taylor2004, McKane2005,Traulsen2005, Reichenbach2006, Melbinger2010}, and mobility of individuals~\cite{Nowak1992, Durrett1994, Frachebourg1996, Mobilia2006, Reichenbach2007a, Abta2008, Rulands2011}. 
An intriguing question is how the stability of ecosystems depends on the interaction network between species. Is it the topology of the network (whose links may arise through predation, competition over common resources, or mutual cooperation) that sets the level of biodiversity? And how important is the strength of a single interaction link? 
Stable coexistence can, for example, be observed for natural populations in non-hierarchical networks that are comprised of species that interact in a competitive and predator-prey like manner ~\cite{Buss1979,Sinervo1996}. 
By understanding the interplay between the structure of the interaction network and the strengths of its links, it is possible to reveal mechanisms that underlie this stability.

\begin{figure}[b!]
\centering
\includegraphics[width=\columnwidth]{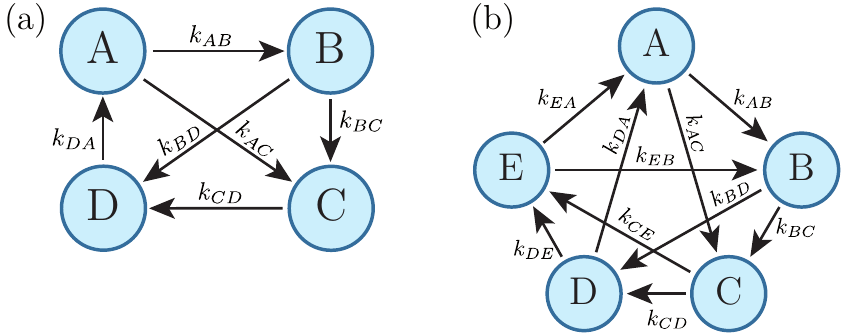}
\caption{(Color online) Two interaction topologies specifying the conservative LV systems. (a)~The general cyclic four species systems (4SS). (b)~The general cyclic five species system (5SS) as a natural extension of the rock-papers-scissors configuration (RPS)~\cite{BBT2008}.
\label{fig:topologies}}
\end{figure}

A paradigm in addressing these ecologically important questions from a theoretical perspective are Lotka-Volterra (LV) models~\cite{Lotka1920, Volterra1931} in which the total biomass of species is conserved.
These conservative LV systems~\cite{Goel1971, Itoh1971, Frachebourg1996} originate in the well-mixed limit from agent-based formulations of reaction-diffusion systems, where individuals of $S$ different species $A_1, A_2, \dots, A_S$ compete directly with each other following the simplified reaction scheme~\cite{Tainaka1989}: $A_i + A_j \longrightarrow A_i+A_i$. Species $A_i$ beats species $A_j$ with rate $k_{ij}$ and immediately replaces an individual of species $A_j$ with an own offspring. Species $A_j$ is thus degraded at the same rate such that the interaction matrix $G_S = \{k_{i j}\}_{i,j}$ is \emph{skew-symmetric}. 
The interaction network can be visualized by a graph; see Fig.~\ref{fig:topologies}.
Neglecting demographic fluctuations~\cite{Dobrinevski2012}, the deterministic dynamics for the species' \emph{concentration} vector $\vec{x} = (x_1,\dots, x_S)^T$ is given by the rate equations (REs):
\begin{align}
\partial_t x_i &=x_i\cdot(G_S\vec{x})_i\ ,\quad \text{for all } i = 1, \dots, S\ .
\label{eq:rate_equations}
\end{align}

This conservative LV model has been investigated as a prototype to understand principles of biodiversity from a theoretical point of view~\cite{Frean2001, Reichenbach2006}.  While these systems are also of central importance to many other fields of science (e.g., plasma physics~\cite{Zakharov1974,*Manakov1975}, evolutionary game theory~\cite{Hofbauer1998, Nowak}, and chemical kinetics~\cite{VanKampen2007}), no general scheme to classify coexistence, survival, and extinction of species has been established so far.
It is frequently assumed that the topology of the interaction network alone determines coexistence of species~\cite{Laird2009, Li2011}, i.e., that such systems can be regarded as Boolean networks~\cite{Klemm2005}. Recent investigations of specific topologies indicate, however, that knowledge about the network topology may not suffice to conclude whether all species coexist or if some of them go extinct~\cite{Case2010, Durney2011,Durney2012}.
These questions on global stability properties have been previously addressed successfully for various particular LV systems~\cite{Goh1977,*Redheffer1984, *Zeeman1995, *Takeuchi, Hofbauer1998} and for hierarchical networks~\cite{Chauvet2002, Cheon2003}.

In this letter, we present a general classification of coexistence scenarios in conservative LV~networks with an arbitrary number of species. 
We elucidate the consequences of the interplay between the network structure and the strengths of its interaction links on global stability. 
By analyzing conserved quantities, we find conditions on the reaction rates that yield coexistence of all species.
In our mathematical framework this amounts to the characterization of positive kernel elements of the interaction matrix: By employing the algebraic concept of the Pfaffian of a skew-symmetric matrix, we are able to generalize previous approaches~\cite{Zia2011, Durney2011} and to quantify the extinction process when no conserved quantities exist. 
We illustrate our general results for coexistence and survival scenarios  of four and five species systems (4SS and 5SS), cf.~Fig.~\ref{fig:topologies}.
Moreover, we demonstrate the implications of our findings for the stability of stochastic systems: We show how the extinction time diverges with the distance to the critical rate at which coexistence of all species is observed.

%%%%%%%%%%%%%%%%
% General properties of the rate equations
%%%%%%%%%%%%%%%%

First, we discuss some general results for the REs~\eqref{eq:rate_equations} before the specific interaction topologies in Fig.~\ref{fig:topologies} are analyzed. 
In order to characterize the stability of the generic LV system, we study conserved quantities.
We elaborate on the form of conserved quantities, under which conditions they exist at all, and how many conserved quantities there are for a given interaction network.
Since the interaction matrix $G_S$ is skew-symmetric, the REs~\eqref{eq:rate_equations} conserve the sum over all species' concentrations $\tau_0 = x_1+\dots x_S$, independent of the interaction scheme. Hence, the dynamics can be normalized onto the 
($S-1$)-dimensional simplex where all concentrations are non-negative and add up to 1. The vertices  of the simplex correspond to the extinction of all but one species, its edges reflect the extinction of all but two species, and so on.
Further conserved quantities have previously been derived as $\tau = x_1^{p_1} \dots  x_S^{p_S}$~\cite{Volterra1931, Hofbauer1998,Zia2011}. Interestingly, these conserved quantities can be obtained from solutions of the linear problem $G_S\vec{p} = \vec{0}$ because $\dot{\tau} = -\tau\mean{G_S\vec{p},\vec{x}}$, with $\vec{p} = (p_1, \dots, p_S)^T$.
One infers that $\tau$ is conserved if the exponent vector $\vec{p}$ is an eigenvector corresponding to eigenvalue 0~\cite{Zia2011}, or in other words, if $\vec{p}$ lies in the kernel of the matrix $G_S$.

Coexistence means that all concentrations stay away from the boundary of the simplex by a finite distance for all times. 
Since the species' concentrations are bounded to the interval $[0,1]$, one concludes from the structure of the conserved quantity $\tau$ that \emph{all} $S$ species coexist if the kernel of the interaction matrix is positive, i.e., one finds an element $\vec{p}$ in the kernel of $G_S$ with positive entries $p_i>0$ for all $i$.
Hence, to reveal coexistence scenarios in the conservative LV model, one has to characterize the kernel of the interaction matrix $G_S$ and  identify its positive elements. 
Note that this conclusion goes beyond stating that a positive kernel element corresponds to a stationary point in the inside of the simplex; see REs~\eqref{eq:rate_equations}.

\begin{figure}[t!]%
\centering
\includegraphics[width=\columnwidth]{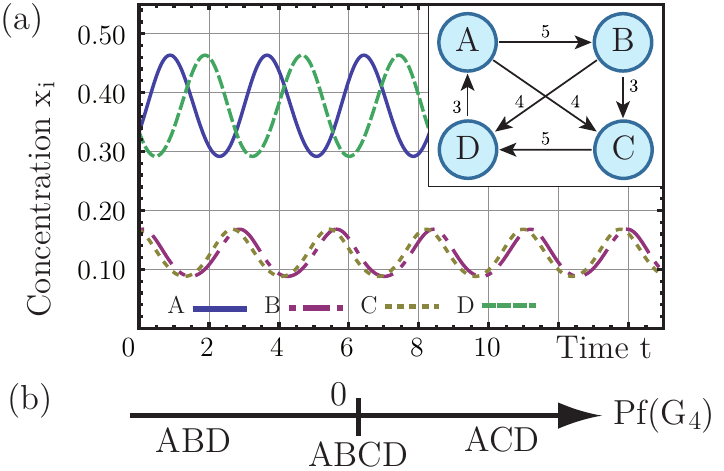}
\caption{(Color online) Coexistence and survival in the general cyclic 4SS are controlled by the Pfaffian of the interaction matrix. (a)~For $\Pf{G_4} = 0$, one obtains coexistence of all species on periodic orbits. (b)~Deterministic survival diagram: for $\Pf{G_4}<0$, species $A, B, $ and $D$ survive in a stable RPS configuration, whereas $A, C, $ and $D$ survive for $\Pf{G_4}>0$.
\label{fig:4SS}}
\end{figure}

The existence of conserved quantities constrains the dynamics to a submanifold of the simplex whose dimension $D_c$ is determined as follows. The rank of a skew-symmetric matrix is always even, because its non-zero eigenvalues are purely imaginary, conjugate pairs. The rank-nullity theorem~\cite{Meyer2000} then implies that the dimension of the kernel of $G_S$ is odd whenever $S$ is odd, and even whenever $S$ is even. Each linearly independent kernel element gives rise to an independent conserved quantity $\tau$ which constrains the degrees of freedom of the trajectory. Together with $\tau_0$, one finds that the dynamics in case of non-stationary motion is constrained to a deformed sphere of dimension $D_c = S-1-\dim\Ker{G_S}$ for a positive kernel; see the Supplemental Material (SM) for mathematical details. Thus, coexistence in high-dimensional systems is generically observed on non-periodic trajectories ($D_c>1$); see Movie~M1 of SM. Only if the reaction rates are fine-tuned to a positive and maximal kernel of dimension $S-2$, the dynamics is restricted to periodic orbits ($D_c=1$); see Fig.~\ref{fig:4SS}(a) and Movie M2 of SM. In particular, for $S=3$ or 4, a positive kernel immediately implies coexistence on periodic orbits. This follows from the fact that with three species, the kernel is always one-dimensional. 
For the general 4SS, the dimension of the kernel of the interaction matrix is either 0 or 2.  A two-dimensional, positive kernel yields coexistence on periodic orbits; see Fig.~\ref{fig:4SS}(a). If $\dim\Ker{G_S} = 0$, i.e., if the kernel is trivial, one observes extinction of species as detailed below.

Next, we focus on the mapping between the reaction rates in $G_S$ and its kernel elements in order to find the stationary points. 
To this end, we apply the concepts of the Pfaffian and of the adjugate matrix~\cite{Cullis1913, Meyer2000}. 
The Pfaffian is a simpler form of the determinant tailored to skew-symmetric matrices with the property that its square equals the value of the determinant. In contrast to the non-negative determinant of skew-symmetric matrices, the Pfaffian carries a sign which will turn out to be crucial for our purposes.  For a skew-symmetric matrix, the Pfaffian can be computed recursively as:
\begin{align}
\Pf{G_S} =\sum_{i = 2}^{S}(-1)^i \cdot k_{1i}\cdot\Pf{G_{\hat{1}\hat{i}}}\ ,
\label{eq:Pfaffian}
\end{align}
with $G_{\hat{1}\hat{i}}$ being the matrix where both the first and $i$-th column and row have been removed from the matrix $G_S$.
The Pfaffian of a $2\times2$ skew-symmetric matrix $G_2 = \{k_{AB}\}$,
%%
%\begin{align}\nonumber
%G_2 =  \left( \begin{array}{cc}
%0 & k_{AB} \\
%-k_{AB} &0 \end{array} \right)\ ,
%\end{align}
%%
is given by $\Pf{G_2} =k_{AB}$. For the interaction matrix corresponding to the LV network in Fig.~\ref{fig:topologies}(a),
\begin{align}\nonumber
G_4 =  \left( \begin{array}{cccc}
0 & k_{AB} & k_{AC} & -k_{DA} \\
-k_{AB} &0 & k_{BC}& k_{BD}\\
-k_{AC} & -k_{BC}& 0& k_{CD}\\
k_{DA} & -k_{BD}& -k_{CD}& 0\end{array} \right)\ ,
%\label{eq:4SS_matrix}
\end{align}
the Pfaffian is  $\Pf{G_4}= k_{AB}k_{CD}-k_{AC}k_{BD}-k_{DA}k_{BC}$. 

The Pfaffian always vanishes for odd $S$ as opposed to systems with an even number of species~\cite{Cullis1913}. In the latter case, the Pfaffian is zero only if a constraint on the reaction rates is fulfilled. If the Pfaffian vanishes, one finds more kernel elements than just the null vector and, thus, conserved quantities of form $\tau$ exist. 
In the following, we distinguish between even and odd $S$.

For an even number of species and a two-dimensional kernel, positive kernel elements can be identified via the adjugate matrix $R_S$ which is a generalized inverse of the interaction matrix such that $G_S\cdot R_S = -\Pf{G_S}\cdot \mathbb{I}_{S}$, with $\mathbb{I}_{S}$ being the identity matrix~\cite{Cullis1913}. 
The adjugate matrix can be computed as $(R_S)_{ij} = (-1)^\sigma\mathrm{Pf}\big(G_{\hat{i}\hat{j}}\big)$ where $(-1)^\sigma$ denotes the sign of the permutation $\sigma = (i\ j\ 1\dots \hat{i}\dots \hat{j}\dots S)$, and the columns of $R_S$ give two independent kernel elements of $G_S$. 

As an example, consider again the general cyclic 4SS depicted in Fig.~\ref{fig:topologies}(a). By setting all reaction rates equal to each other (e.g., to 1), the Pfaffian does not vanish and, therefore, not all species can coexist. Only when the rates are chosen such that $\Pf{G_4} = 0$, do we obtain two independent kernel elements  of $G_4$: From its adjugate matrix, $R_4$,
%%
%\begin{align}\nonumber
%R_4 =  \left( \begin{array}{cccc}
%0 & k_{CD} & -k_{BD} & k_{BC} \\
%-k_{CD} & 0 & -k_{DA} & -k_{AC}\\
%k_{BD} & k_{DA} & 0 & k_{AB}\\
%-k_{BC} & k_{AC} & -k_{AB} & 0\end{array} \right)\ ,
%%\label{eq:4SS_reciprocal}
%\end{align}
%%
we identify $\vec{p}_1= (k_{CD}, 0, k_{DA}, k_{AC})^T$ and $\vec{p}_2= (k_{BD}, k_{DA}, 0,k_{AB})^T$. We infer the two conserved quantities $\tau_1 = x_A^{k_{CD}}x_C^{k_{DA}}x_D^{k_{AC}}$ and $\tau_2 = x_A^{k_{BD}}x_B^{k_{DA}}x_D^{k_{AB}}$, and conclude that the kernel is positive and coexistence occurs on periodic orbits; see Fig.~\ref{fig:4SS}(a). 
Hence, classifying LV networks in terms of their topology is incomplete; the strengths of the interaction links are crucial in general. 

In general, if the Pfaffian for a system with even $S$ is non-zero, i.e., when only the null vector lies in the kernel, coexistence of all species is not possible. Still one can quantify the extinction process by generalizing an approach of Durney et al.~\cite{Durney2011} for a system with $S=4$ to systems composed of an arbitrary even number of species. We define the function $\rho = x_1^{q_1}\dots x_S^{q_S}$ in the same way as the conserved quantity $\tau$, but this time choosing the exponent vector $\vec{q}_S = -R_S \vec{1}$ with $\vec{1} = (1, \dots, 1)^T$.
It is straightforward to show that this function evolves exponentially in time: 
\begin{align}
\rho(t) = \rho(0)\cdot e^{-\Pf{G_S}\cdot t} \ ,
\label{eq:rho_evolution}
\end{align}
generalizing previous investigations \cite{Case2010, Durney2011, Durney2012, Dobrinevski2012}. 
It is quite remarkable that $\rho$ quantifies the global collective dynamics of systems with an arbitrary interaction topology and even $S$. Depending on the sign of the Pfaffian, $\rho$ grows or decays exponentially fast with the Pfaffian of the interaction matrix as rate. 
Since the system's dynamics is driven towards the boundary of the simplex, one can conclude on the extinction of some species. This feature of $\rho$ is reminiscent of a Lyapunov function; note also that $\rho$ becomes a conserved quantity $\tau$ if the Pfaffian is zero. An interesting question for future investigations is to ask whether further quantities exist that characterize the dynamics of conservative LV networks.

For the  general 4SS shown in Fig.~\ref{fig:topologies}(a), we find $\vec{q}_4  = (-k_{CD}+k_{BD}-k_{BC}, k_{CD}+k_{DA}+k_{AC}, -k_{BD}-k_{DA}-k_{AB}, k_{BC}-k_{AC}+k_{AB})^T$. The fact that $(\vec{q}_4)_2$ is always positive suggests that species $B$ goes extinct for a positive Pfaffian, and that the converse holds true for $(\vec{q}_4)_3$ and species $C$ for a negative Pfaffian. In both cases, the system tends to a stable rock-paper-scissors (RPS) configuration. In summary, we derive the survival diagram shown in Fig.~\ref{fig:4SS}(b). Interestingly, $A$ and $D$ always survive in this topology although $D$ can be easily tuned to be the weakest species.
We emphasize that this result depends on the sign of the Pfaffian and cannot be obtained from applying the concept of the determinant.
Again, since the Pfaffian of the interaction matrix characterizes the dynamics of this 4SS, its topology alone does not determine the long-time dynamics.
 These findings unify previous results for other 4SS ~\cite{Case2010, Durney2011, Dobrinevski2012}, and show that rules like ``survival of the strongest'' or ``survival of the weakest''~\cite{Frean2001, Berr2009} cannot be formulated in general. 

For an odd number of  species, the kernel of $G_S$ is always nontrivial. 
In general, if $\dim\ker{G_S} = 1$, we determine the independent  kernel element via the adjugate vector~\cite{Cullis1913},
$\vec{r}_S = \left(\Pf{G_{\hat{1}}}, -\Pf{G_{\hat{2}}}, \dots, \Pf{G_{\hat{S}}}\right)^T$, 
which enables us to investigate the influence of the reaction rates on the survival scenarios.
For $S=3$, only the well-studied RPS topology~\cite{Hofbauer1998, Reichenbach2006} leads to a positive adjugate vector $\vec{r}_3$. 
In other words, coexistence of all three species depends only on the topology of the network. This behavior is unique to $S=3$ and changes dramatically for systems with more than three species. 

\begin{figure}
\centering
\includegraphics[width=\columnwidth]{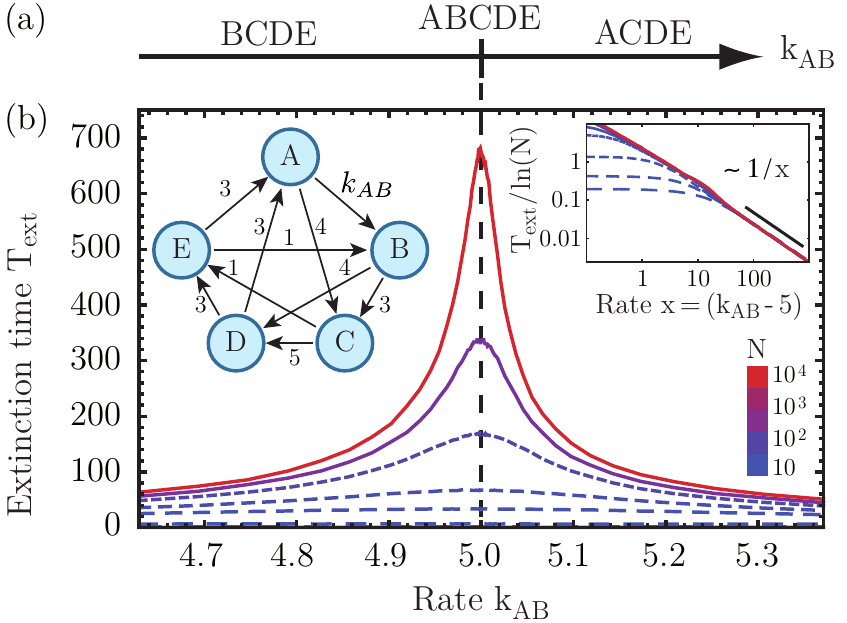}
\caption{(Color online) Stability of the cyclic 5SS. (a) For the interaction scheme (left inset of (b)), one obtains coexistence of all species for the critical rate $k_{AB} = 5$. (b) Stability of the stochastic system, reflected by the extinction time $T_\mathrm{ext}$, peaks at the critical rate, which becomes more pronounced as $N\rightarrow\infty$. We find a scaling law for $T_\mathrm{ext}$ in the distance to the critical rate (right inset). Initial conditions were chosen as $\vec{x}(0) = 1/5\cdot\vec{1}$. Larger line gap corresponds to smaller $N$.
\label{fig:5SS_stochastic}}
\end{figure}

We illustrate the importance of the reaction rates for a system of five interacting species; see Fig.~\ref{fig:topologies}(b). This interaction topology where each species dominates two species and is outperformed by the two remaining species, recently gained attention as a natural extension of the RPS game~\cite{BBT2008, Laird2009, Hawick2011}.  
For specificity, we investigate the dependence of the survival scenarios on the rate $k_{AB}$ with which species $A$ beats species $B$ and chose the other rates (see Fig.~\ref{fig:5SS_stochastic}(b), left inset) such that either five or four species survive depending on the value of $k_{AB}$; see Fig.~\ref{fig:5SS_stochastic}(a). The kernel of the interaction matrix depends on  $k_{AB}$ and is characterized by the adjugate vector $\vec{r}_5 = (0, 0, 3 k_{AB} - 15, 5-k_{AB}, 5 k_{AB}-25)^T$. For $k_{AB}\neq 5$, the kernel is one-dimensional and non-positive, and four species survive. In contrast, for $k_{AB}= 5$, $\vec{r}_5$ equals the null vector which in turn means that the kernel becomes three-dimensional~\cite{Cullis1913}. Since we have ensured that the kernel is also positive, we obtain coexistence of all five species on periodic orbits ($D_c = 1$).
%

%%%%%%%%%%%%%%%%
% Stochastics
%%%%%%%%%%%%%%%%
Finally, we discuss the implications of our findings by asking how demographic noise affects the stability of stochastic LV systems. We analyze ecological LV systems with a finite number $N$ of interacting individuals in the eye of the knowledge gained from the deterministic analysis. It has been shown that due to demographic fluctuations the system ultimately reaches an absorbing state that is characterized by the extinction of all but one species~\cite{Krapivsky2010,Frachebourg1998, Antal2006, Parker2009}. Moreover, the scaling behavior of the mean extinction time with the system size $N$ characterizes the stability of the interaction network~\cite{Antal2006, Reichenbach2007a}. 

As an example, we continue the discussion of the 5SS from Fig.~\ref{fig:5SS_stochastic}(b), left inset. We have carried out extensive computer simulations employing the Gillespie algorithm~\cite{Gillespie1976} to measure the time $T_\mathrm{ext}$ until the first species has become extinct for different system sizes $N$ and different reaction rates $k_{AB}$. The results are displayed in Fig.~\ref{fig:5SS_stochastic}(b) and highlight the significance of the deterministic drift underlying the stochastic dynamics. We observe a peak in the extinction time as the reaction rate $k_{AB}$ approaches the critical value $k_\mathrm{cr} = 5$ for which we obtain coexistence of all species in the deterministic case. The divergence of the extinction time for $k_{AB}\rightarrow k_\mathrm{cr}$ becomes more pronounced for larger system sizes as the system reaches the deterministic limit for $N\rightarrow\infty$.

A scaling analysis reveals how the extinction time peaks in the vicinity of the coexistence scenario. Near the critical rate, the extinction time scales linearly with the system size leading to neutrally stable interaction networks~\cite{Reichenbach2006, Cremer2009, Dobrinevski2012}. At larger distance from the critical rate, the deterministic driving force to the absorbing boundary becomes more dominant than the demographic fluctuations; see Fig.~\ref{fig:5SS_stochastic}(b), right inset. 
The interplay between the stochastic and deterministic forces is reflected by the scaling law:
\begin{align}\label{eq:scaling_law}
T_\mathrm{ext}\propto 
\begin{cases}
N &\mbox{for } k_{AB} = k_\mathrm{cr}\ ,\\
\frac{\ln{N}}{|k_{AB}-k_\mathrm{cr}|} &\mbox{for } k_{AB} \neq k_\mathrm{cr}\ ,
\end{cases}
\end{align}
which extends the linear scaling  $T_\mathrm{ext}\propto N$ of neutral coexistence. We observe a power-law dependence in the distance of the reaction rates to the critical rate and logarithmic scaling with $N$ for attracting boundaries~\cite{Reichenbach2006, Frey2010}. 

The observed scaling law~\eqref{eq:scaling_law} for $k_{AB} \neq k_\mathrm{cr}$ can be attributed to the exponentially fast extinction of species $x_i=x_i(0) \exp{(-\alpha_i t)}$; see Eq.~\eqref{eq:rate_equations}. The extinction rate $\alpha_i$ is computed via the temporal average over the trajectory $\mean{\vec{x}}$ as $\alpha_i = -(G_S\mean{\vec{x}})_i$, which becomes linear in the distance to the critical rate $|k_{AB}-k_\mathrm{cr}|$ for large times. The logarithmic dependence on $N$ follows by defining that a species with concentration $x_i$ less than $1/N$ has become extinct.
With this scaling behavior at hand, we are able to compare the ecological stability of different interaction networks based on our  analysis of the REs~\eqref{eq:rate_equations}. 

%%%%%%%%%%%%%%%%
% Conclusion
%%%%%%%%%%%%%%%%
In this Letter, we investigated global stability properties of conservative LV networks. By employing the Pfaffian of the interaction matrix, we revealed the relation between the reaction rates and the conditions for coexistence, and exemplified the implications  for the stability of ecological networks with finite populations.
We expect that our results will also stimulate further progress for the investigation of extinction scenarios. Beyond analyzing whether an ecosystem is stable or unstable, it would be highly interesting to actually predict \emph{which} of its species ultimately survive for a general conservative LV system. 
This would, for example, allow us to predict the eventual outcome of an unstable version of the five species system shown in Fig.~\ref{fig:topologies}(b), and to formulate the conditions under which 3- or 4-species cycles are attained. First insights into these extinction dynamics will be outlined in a future publication~\cite{Knebel2013}.
We believe that a full characterization of general conservative LV dynamics is possible.

\clearpage
\newpage
\begin{acknowledgments}
We would like to thank Steffen Rulands, Lukas Darnst\"adt, Falk T\"oppel, and Michal Oszmaniec for helpful discussions. This project was supported by the Deutsche Forschungsgemeinschaft in the framework of the SFB~TR~12 ``Symmetry and Universality in Mesoscopic Systems", and the German Excellence Initiative via the program ``Nanosystems Initiative Munich'' (NIM). J.K. acknowledges funding by the Studienstiftung des Deutschen Volkes.  
\end{acknowledgments}


\begin{thebibliography}{56}%
\makeatletter
\providecommand \@ifxundefined [1]{%
 \@ifx{#1\undefined}
}%
\providecommand \@ifnum [1]{%
 \ifnum #1\expandafter \@firstoftwo
 \else \expandafter \@secondoftwo
 \fi
}%
\providecommand \@ifx [1]{%
 \ifx #1\expandafter \@firstoftwo
 \else \expandafter \@secondoftwo
 \fi
}%
\providecommand \natexlab [1]{#1}%
\providecommand \enquote  [1]{``#1''}%
\providecommand \bibnamefont  [1]{#1}%
\providecommand \bibfnamefont [1]{#1}%
\providecommand \citenamefont [1]{#1}%
\providecommand \href@noop [0]{\@secondoftwo}%
\providecommand \href [0]{\begingroup \@sanitize@url \@href}%
\providecommand \@href[1]{\@@startlink{#1}\@@href}%
\providecommand \@@href[1]{\endgroup#1\@@endlink}%
\providecommand \@sanitize@url [0]{\catcode `\\12\catcode `\$12\catcode
  `\&12\catcode `\#12\catcode `\^12\catcode `\_12\catcode `\%12\relax}%
\providecommand \@@startlink[1]{}%
\providecommand \@@endlink[0]{}%
\providecommand \url  [0]{\begingroup\@sanitize@url \@url }%
\providecommand \@url [1]{\endgroup\@href {#1}{\urlprefix }}%
\providecommand \urlprefix  [0]{URL }%
\providecommand \Eprint [0]{\href }%
\providecommand \doibase [0]{http://dx.doi.org/}%
\providecommand \selectlanguage [0]{\@gobble}%
\providecommand \bibinfo  [0]{\@secondoftwo}%
\providecommand \bibfield  [0]{\@secondoftwo}%
\providecommand \translation [1]{[#1]}%
\providecommand \BibitemOpen [0]{}%
\providecommand \bibitemStop [0]{}%
\providecommand \bibitemNoStop [0]{.\EOS\space}%
\providecommand \EOS [0]{\spacefactor3000\relax}%
\providecommand \BibitemShut  [1]{\csname bibitem#1\endcsname}%
\let\auto@bib@innerbib\@empty
%</preamble>
\bibitem [{\citenamefont {May}(1973)}]{May}%
  \BibitemOpen
  \bibfield  {author} {\bibinfo {author} {\bibfnamefont {R.~M.}\ \bibnamefont
  {May}},\ }\href@noop {} {\emph {\bibinfo {title} {{Stability and complexity
  in model ecosystems}}}}\ (\bibinfo  {publisher} {Princeton University
  Press},\ \bibinfo {address} {Princeton, NJ},\ \bibinfo {year}
  {1973})\BibitemShut {NoStop}%
\bibitem [{\citenamefont {Montoya}\ \emph {et~al.}(2006)\citenamefont
  {Montoya}, \citenamefont {Pimm},\ and\ \citenamefont
  {Sol\'{e}}}]{Montoya2006}%
  \BibitemOpen
  \bibfield  {author} {\bibinfo {author} {\bibfnamefont {J.~M.}\ \bibnamefont
  {Montoya}}, \bibinfo {author} {\bibfnamefont {S.~L.}\ \bibnamefont {Pimm}}, \
  and\ \bibinfo {author} {\bibfnamefont {R.~V.}\ \bibnamefont {Sol\'{e}}},\
  }\href {\doibase 10.1038/nature04927} {\bibfield  {journal} {\bibinfo
  {journal} {Nature}\ }\textbf {\bibinfo {volume} {442}},\ \bibinfo {pages}
  {259} (\bibinfo {year} {2006})}\BibitemShut {NoStop}%
\bibitem [{\citenamefont {Szab\'{o}}\ and\ \citenamefont
  {F\'{a}th}(2007)}]{Szabo2007b}%
  \BibitemOpen
  \bibfield  {author} {\bibinfo {author} {\bibfnamefont {G.}~\bibnamefont
  {Szab\'{o}}}\ and\ \bibinfo {author} {\bibfnamefont {G.}~\bibnamefont
  {F\'{a}th}},\ }\href {\doibase doi:10.1016/j.physrep.2007.04.004} {\bibfield
  {journal} {\bibinfo  {journal} {Phys. Rep.}\ }\textbf {\bibinfo {volume}
  {446}},\ \bibinfo {pages} {97} (\bibinfo {year} {2007})}\BibitemShut
  {NoStop}%
\bibitem [{\citenamefont {Mathiesen}\ \emph {et~al.}(2011)\citenamefont
  {Mathiesen}, \citenamefont {Mitarai}, \citenamefont {Sneppen},\ and\
  \citenamefont {Trusina}}]{Mathiesen2011}%
  \BibitemOpen
  \bibfield  {author} {\bibinfo {author} {\bibfnamefont {J.}~\bibnamefont
  {Mathiesen}}, \bibinfo {author} {\bibfnamefont {N.}~\bibnamefont {Mitarai}},
  \bibinfo {author} {\bibfnamefont {K.}~\bibnamefont {Sneppen}}, \ and\
  \bibinfo {author} {\bibfnamefont {A.}~\bibnamefont {Trusina}},\ }\href
  {\doibase 10.1103/PhysRevLett.107.188101} {\bibfield  {journal} {\bibinfo
  {journal} {Phys. Rev. Lett.}\ }\textbf {\bibinfo {volume} {107}},\ \bibinfo
  {pages} {188101} (\bibinfo {year} {2011})}\BibitemShut {NoStop}%
\bibitem [{\citenamefont {Taylor}\ \emph {et~al.}(2004)\citenamefont {Taylor},
  \citenamefont {Fudenberg}, \citenamefont {Sasaki},\ and\ \citenamefont
  {Nowak}}]{Taylor2004}%
  \BibitemOpen
  \bibfield  {author} {\bibinfo {author} {\bibfnamefont {C.}~\bibnamefont
  {Taylor}}, \bibinfo {author} {\bibfnamefont {D.}~\bibnamefont {Fudenberg}},
  \bibinfo {author} {\bibfnamefont {A.}~\bibnamefont {Sasaki}}, \ and\ \bibinfo
  {author} {\bibfnamefont {M.}~\bibnamefont {Nowak}},\ }\href {\doibase
  10.1016/j.bulm.2004.03.004} {\bibfield  {journal} {\bibinfo  {journal} {Bull.
  Math. Biol.}\ }\textbf {\bibinfo {volume} {66}},\ \bibinfo {pages} {1621}
  (\bibinfo {year} {2004})}\BibitemShut {NoStop}%
\bibitem [{\citenamefont {McKane}\ and\ \citenamefont
  {Newman}(2005)}]{McKane2005}%
  \BibitemOpen
  \bibfield  {author} {\bibinfo {author} {\bibfnamefont {A.~J.}\ \bibnamefont
  {McKane}}\ and\ \bibinfo {author} {\bibfnamefont {T.~J.}\ \bibnamefont
  {Newman}},\ }\href {\doibase 10.1103/PhysRevLett.94.218102} {\bibfield
  {journal} {\bibinfo  {journal} {Phys. Rev. Lett.}\ }\textbf {\bibinfo
  {volume} {94}},\ \bibinfo {pages} {218102} (\bibinfo {year}
  {2005})}\BibitemShut {NoStop}%
\bibitem [{\citenamefont {Traulsen}\ \emph {et~al.}(2005)\citenamefont
  {Traulsen}, \citenamefont {Claussen},\ and\ \citenamefont
  {Hauert}}]{Traulsen2005}%
  \BibitemOpen
  \bibfield  {author} {\bibinfo {author} {\bibfnamefont {A.}~\bibnamefont
  {Traulsen}}, \bibinfo {author} {\bibfnamefont {J.~C.}\ \bibnamefont
  {Claussen}}, \ and\ \bibinfo {author} {\bibfnamefont {C.}~\bibnamefont
  {Hauert}},\ }\href {\doibase 10.1103/PhysRevLett.95.238701} {\bibfield
  {journal} {\bibinfo  {journal} {Phys. Rev. Lett.}\ }\textbf {\bibinfo
  {volume} {95}},\ \bibinfo {pages} {238701} (\bibinfo {year}
  {2005})}\BibitemShut {NoStop}%
\bibitem [{\citenamefont {Reichenbach}\ \emph {et~al.}(2006)\citenamefont
  {Reichenbach}, \citenamefont {Mobilia},\ and\ \citenamefont
  {Frey}}]{Reichenbach2006}%
  \BibitemOpen
  \bibfield  {author} {\bibinfo {author} {\bibfnamefont {T.}~\bibnamefont
  {Reichenbach}}, \bibinfo {author} {\bibfnamefont {M.}~\bibnamefont
  {Mobilia}}, \ and\ \bibinfo {author} {\bibfnamefont {E.}~\bibnamefont
  {Frey}},\ }\href {\doibase 10.1103/PhysRevE.74.051907} {\bibfield  {journal}
  {\bibinfo  {journal} {Phys. Rev. E}\ }\textbf {\bibinfo {volume} {74}},\
  \bibinfo {pages} {051907} (\bibinfo {year} {2006})}\BibitemShut {NoStop}%
\bibitem [{\citenamefont {Melbinger}\ \emph {et~al.}(2010)\citenamefont
  {Melbinger}, \citenamefont {Cremer},\ and\ \citenamefont
  {Frey}}]{Melbinger2010}%
  \BibitemOpen
  \bibfield  {author} {\bibinfo {author} {\bibfnamefont {A.}~\bibnamefont
  {Melbinger}}, \bibinfo {author} {\bibfnamefont {J.}~\bibnamefont {Cremer}}, \
  and\ \bibinfo {author} {\bibfnamefont {E.}~\bibnamefont {Frey}},\ }\href
  {\doibase 10.1103/PhysRevLett.105.178101} {\bibfield  {journal} {\bibinfo
  {journal} {Phys. Rev. Lett.}\ }\textbf {\bibinfo {volume} {105}},\ \bibinfo
  {pages} {178101} (\bibinfo {year} {2010})}\BibitemShut {NoStop}%
\bibitem [{\citenamefont {Nowak}\ and\ \citenamefont {May}(1992)}]{Nowak1992}%
  \BibitemOpen
  \bibfield  {author} {\bibinfo {author} {\bibfnamefont {M.}~\bibnamefont
  {Nowak}}\ and\ \bibinfo {author} {\bibfnamefont {R.}~\bibnamefont {May}},\
  }\href
  {http://www.ped.fas.harvard.edu/people/faculty/publications\_nowak/Nature92.pdf}
  {\bibfield  {journal} {\bibinfo  {journal} {Nature}\ }\textbf {\bibinfo
  {volume} {359}},\ \bibinfo {pages} {826} (\bibinfo {year}
  {1992})}\BibitemShut {NoStop}%
\bibitem [{\citenamefont {Durrett}\ and\ \citenamefont
  {Levin}(1994)}]{Durrett1994}%
  \BibitemOpen
  \bibfield  {author} {\bibinfo {author} {\bibfnamefont {R.}~\bibnamefont
  {Durrett}}\ and\ \bibinfo {author} {\bibfnamefont {S.}~\bibnamefont
  {Levin}},\ }\href {http://en.scientificcommons.org/26103908
  http://www.math.umaine.edu/~hiebeler/papers/durrett-levin-94.pdf} {\bibfield
  {journal} {\bibinfo  {journal} {Theor. Popul. Biol.}\ }\textbf {\bibinfo
  {volume} {46}},\ \bibinfo {pages} {363} (\bibinfo {year} {1994})}\BibitemShut
  {NoStop}%
\bibitem [{\citenamefont {Frachebourg}\ \emph {et~al.}(1996)\citenamefont
  {Frachebourg}, \citenamefont {Krapivsky},\ and\ \citenamefont
  {Ben-Naim}}]{Frachebourg1996}%
  \BibitemOpen
  \bibfield  {author} {\bibinfo {author} {\bibfnamefont {L.}~\bibnamefont
  {Frachebourg}}, \bibinfo {author} {\bibfnamefont {P.~L.}\ \bibnamefont
  {Krapivsky}}, \ and\ \bibinfo {author} {\bibfnamefont {E.}~\bibnamefont
  {Ben-Naim}},\ }\href {http://www.ncbi.nlm.nih.gov/pubmed/9965838} {\bibfield
  {journal} {\bibinfo  {journal} {Phys. Rev. E}\ }\textbf {\bibinfo {volume}
  {54}},\ \bibinfo {pages} {6186} (\bibinfo {year} {1996})}\BibitemShut
  {NoStop}%
\bibitem [{\citenamefont {Mobilia}\ \emph {et~al.}(2006)\citenamefont
  {Mobilia}, \citenamefont {Georgiev},\ and\ \citenamefont
  {T\"{a}uber}}]{Mobilia2006}%
  \BibitemOpen
  \bibfield  {author} {\bibinfo {author} {\bibfnamefont {M.}~\bibnamefont
  {Mobilia}}, \bibinfo {author} {\bibfnamefont {I.~T.}\ \bibnamefont
  {Georgiev}}, \ and\ \bibinfo {author} {\bibfnamefont {U.~C.}\ \bibnamefont
  {T\"{a}uber}},\ }\href {\doibase 10.1007/s10955-006-9146-3} {\bibfield
  {journal} {\bibinfo  {journal} {J. Stat. Phys.}\ }\textbf {\bibinfo {volume}
  {128}},\ \bibinfo {pages} {447} (\bibinfo {year} {2006})}\BibitemShut
  {NoStop}%
\bibitem [{\citenamefont {Reichenbach}\ \emph {et~al.}(2007)\citenamefont
  {Reichenbach}, \citenamefont {Mobilia},\ and\ \citenamefont
  {Frey}}]{Reichenbach2007a}%
  \BibitemOpen
  \bibfield  {author} {\bibinfo {author} {\bibfnamefont {T.}~\bibnamefont
  {Reichenbach}}, \bibinfo {author} {\bibfnamefont {M.}~\bibnamefont
  {Mobilia}}, \ and\ \bibinfo {author} {\bibfnamefont {E.}~\bibnamefont
  {Frey}},\ }\href {\doibase 10.1038/nature06095} {\bibfield  {journal}
  {\bibinfo  {journal} {Nature}\ }\textbf {\bibinfo {volume} {448}},\ \bibinfo
  {pages} {1046} (\bibinfo {year} {2007})}\BibitemShut {NoStop}%
\bibitem [{\citenamefont {Abta}\ \emph {et~al.}(2008)\citenamefont {Abta},
  \citenamefont {Schiffer}, \citenamefont {Ben-Ishay},\ and\ \citenamefont
  {Schnerb}}]{Abta2008}%
  \BibitemOpen
  \bibfield  {author} {\bibinfo {author} {\bibfnamefont {R.}~\bibnamefont
  {Abta}}, \bibinfo {author} {\bibfnamefont {M.}~\bibnamefont {Schiffer}},
  \bibinfo {author} {\bibfnamefont {A.}~\bibnamefont {Ben-Ishay}}, \ and\
  \bibinfo {author} {\bibfnamefont {N.~M.}\ \bibnamefont {Schnerb}},\ }\href
  {http://cat.inist.fr/?aModele=afficheN\&cpsidt=20770234} {\bibfield
  {journal} {\bibinfo  {journal} {Theoretical population biology}\ }\textbf
  {\bibinfo {volume} {74}},\ \bibinfo {pages} {273} (\bibinfo {year}
  {2008})}\BibitemShut {NoStop}%
\bibitem [{\citenamefont {Rulands}\ \emph {et~al.}(2011)\citenamefont
  {Rulands}, \citenamefont {Reichenbach},\ and\ \citenamefont
  {Frey}}]{Rulands2011}%
  \BibitemOpen
  \bibfield  {author} {\bibinfo {author} {\bibfnamefont {S.}~\bibnamefont
  {Rulands}}, \bibinfo {author} {\bibfnamefont {T.}~\bibnamefont
  {Reichenbach}}, \ and\ \bibinfo {author} {\bibfnamefont {E.}~\bibnamefont
  {Frey}},\ }\href {\doibase 10.1088/1742-5468/2011/01/L01003} {\bibfield
  {journal} {\bibinfo  {journal} {J. Stat. Phys.}\ }\textbf {\bibinfo {volume}
  {2011}},\ \bibinfo {pages} {L01003} (\bibinfo {year} {2011})}\BibitemShut
  {NoStop}%
\bibitem [{\citenamefont {Buss}\ and\ \citenamefont
  {Jackson}(1979)}]{Buss1979}%
  \BibitemOpen
  \bibfield  {author} {\bibinfo {author} {\bibfnamefont {L.~W.}\ \bibnamefont
  {Buss}}\ and\ \bibinfo {author} {\bibfnamefont {J.~B.~C.}\ \bibnamefont
  {Jackson}},\ }\href@noop {} {\bibfield  {journal} {\bibinfo  {journal} {Am.
  Nat.}\ }\textbf {\bibinfo {volume} {113}},\ \bibinfo {pages} {223} (\bibinfo
  {year} {1979})}\BibitemShut {NoStop}%
\bibitem [{\citenamefont {Sinervo}\ and\ \citenamefont
  {Lively}(1996)}]{Sinervo1996}%
  \BibitemOpen
  \bibfield  {author} {\bibinfo {author} {\bibfnamefont {B.}~\bibnamefont
  {Sinervo}}\ and\ \bibinfo {author} {\bibfnamefont {C.}~\bibnamefont
  {Lively}},\ }\href
  {http://wolfweb.unr.edu/homepage/vpravosu/BehavEcol/Sinervo\&Lively\_1996Nature.pdf}
  {\bibfield  {journal} {\bibinfo  {journal} {Nature}\ }\textbf {\bibinfo
  {volume} {380}},\ \bibinfo {pages} {240} (\bibinfo {year}
  {1996})}\BibitemShut {NoStop}%
\bibitem [{\citenamefont {Lotka}(1920)}]{Lotka1920}%
  \BibitemOpen
  \bibfield  {author} {\bibinfo {author} {\bibfnamefont {A.}~\bibnamefont
  {Lotka}},\ }\href {\doibase 10.1021/ja01453a010} {\bibfield  {journal}
  {\bibinfo  {journal} {J. Am. Chem. Soc.}\ }\textbf {\bibinfo {volume} {42}},\
  \bibinfo {pages} {1595} (\bibinfo {year} {1920})}\BibitemShut {NoStop}%
\bibitem [{\citenamefont {Volterra}(1931)}]{Volterra1931}%
  \BibitemOpen
  \bibfield  {author} {\bibinfo {author} {\bibfnamefont {V.}~\bibnamefont
  {Volterra}},\ }\href@noop {} {\emph {\bibinfo {title} {{Le{\c c}ons sur la
  Th{\'e}orie Math{\'e}matique de la Lutte pour la Vie}}}},\ \bibinfo {edition}
  {1st}\ ed.\ (\bibinfo  {publisher} {Gauthier-Villars},\ \bibinfo {address}
  {Paris},\ \bibinfo {year} {1931})\BibitemShut {NoStop}%
\bibitem [{\citenamefont {Goel}\ \emph {et~al.}(1971)\citenamefont {Goel},
  \citenamefont {Maitra},\ and\ \citenamefont {Montroll}}]{Goel1971}%
  \BibitemOpen
  \bibfield  {author} {\bibinfo {author} {\bibfnamefont {N.~S.}\ \bibnamefont
  {Goel}}, \bibinfo {author} {\bibfnamefont {S.~C.}\ \bibnamefont {Maitra}}, \
  and\ \bibinfo {author} {\bibfnamefont {E.~W.}\ \bibnamefont {Montroll}},\
  }\href {http://rmp.aps.org/abstract/RMP/v43/i2/p231\_1} {\bibfield  {journal}
  {\bibinfo  {journal} {Rev. Mod. Phys.}\ }\textbf {\bibinfo {volume} {43}}
  (\bibinfo {year} {1971})}\BibitemShut {NoStop}%
\bibitem [{\citenamefont {Itoh}(1971)}]{Itoh1971}%
  \BibitemOpen
  \bibfield  {author} {\bibinfo {author} {\bibfnamefont {Y.}~\bibnamefont
  {Itoh}},\ }\href {http://www.ism.ac.jp/~itoh/pub/1971\_a.pdf} {\bibfield
  {journal} {\bibinfo  {journal} {Proc. Jpn. Acad.}\ }\textbf {\bibinfo
  {volume} {47}},\ \bibinfo {pages} {854} (\bibinfo {year} {1971})}\BibitemShut
  {NoStop}%
\bibitem [{\citenamefont {Tainaka}(1989)}]{Tainaka1989}%
  \BibitemOpen
  \bibfield  {author} {\bibinfo {author} {\bibfnamefont {K.~I.}\ \bibnamefont
  {Tainaka}},\ }\href@noop {} {\bibfield  {journal} {\bibinfo  {journal} {Phys.
  Rev. Lett.}\ }\textbf {\bibinfo {volume} {63}},\ \bibinfo {pages} {2688}
  (\bibinfo {year} {1989})}\BibitemShut {NoStop}%
\bibitem [{\citenamefont {Dobrinevski}\ and\ \citenamefont
  {Frey}(2012)}]{Dobrinevski2012}%
  \BibitemOpen
  \bibfield  {author} {\bibinfo {author} {\bibfnamefont {A.}~\bibnamefont
  {Dobrinevski}}\ and\ \bibinfo {author} {\bibfnamefont {E.}~\bibnamefont
  {Frey}},\ }\href {\doibase 10.1103/PhysRevE.85.051903} {\bibfield  {journal}
  {\bibinfo  {journal} {Phys. Rev. E}\ }\textbf {\bibinfo {volume} {85}},\
  \bibinfo {pages} {051903} (\bibinfo {year} {2012})}\BibitemShut {NoStop}%
\bibitem [{\citenamefont {Frean}\ and\ \citenamefont
  {Abraham}(2001)}]{Frean2001}%
  \BibitemOpen
  \bibfield  {author} {\bibinfo {author} {\bibfnamefont {M.}~\bibnamefont
  {Frean}}\ and\ \bibinfo {author} {\bibfnamefont {E.~R.}\ \bibnamefont
  {Abraham}},\ }\href@noop {} {\bibfield  {journal} {\bibinfo  {journal} {Proc.
  R. Soc. B}\ }\textbf {\bibinfo {volume} {268}},\ \bibinfo {pages} {1323}
  (\bibinfo {year} {2001})}\BibitemShut {NoStop}%
\bibitem [{\citenamefont {Zakharov}\ \emph {et~al.}(1974)\citenamefont
  {Zakharov}, \citenamefont {Musher},\ and\ \citenamefont
  {Rubenchik}}]{Zakharov1974}%
  \BibitemOpen
  \bibfield  {author} {\bibinfo {author} {\bibfnamefont {V.}~\bibnamefont
  {Zakharov}}, \bibinfo {author} {\bibfnamefont {S.}~\bibnamefont {Musher}}, \
  and\ \bibinfo {author} {\bibfnamefont {A.}~\bibnamefont {Rubenchik}},\ }\href
  {http://www.jetpletters.ac.ru/ps/1774/article\_26980.pdf} {\bibfield
  {journal} {\bibinfo  {journal} {JETP Lett.}\ }\textbf {\bibinfo {volume}
  {19}},\ \bibinfo {pages} {151} (\bibinfo {year} {1974})}\BibitemShut
  {NoStop}%
\bibitem [{\citenamefont {Manakov}(1975)}]{Manakov1975}%
  \BibitemOpen
  \bibfield  {author} {\bibinfo {author} {\bibfnamefont {S.}~\bibnamefont
  {Manakov}},\ }\href
  {http://www.ams.org/mathscinet/pdf/389107.pdf?pg1=MR\&s1=52:9938\&loc=fromreflist}
  {\bibfield  {journal} {\bibinfo  {journal} {Sov. Phys. JETP}\ }\textbf
  {\bibinfo {volume} {40}},\ \bibinfo {pages} {269} (\bibinfo {year}
  {1975})}\BibitemShut {NoStop}%
\bibitem [{\citenamefont {Hofbauer}\ and\ \citenamefont
  {Sigmund}(1998)}]{Hofbauer1998}%
  \BibitemOpen
  \bibfield  {author} {\bibinfo {author} {\bibfnamefont {J.}~\bibnamefont
  {Hofbauer}}\ and\ \bibinfo {author} {\bibfnamefont {K.}~\bibnamefont
  {Sigmund}},\ }\href@noop {} {\emph {\bibinfo {title} {{Evolutionary Games and
  Population Dynamics}}}},\ \bibinfo {edition} {1st}\ ed.\ (\bibinfo
  {publisher} {Cambridge University Press},\ \bibinfo {address} {Cambridge},\
  \bibinfo {year} {1998})\BibitemShut {NoStop}%
\bibitem [{\citenamefont {Nowak}(2006)}]{Nowak}%
  \BibitemOpen
  \bibfield  {author} {\bibinfo {author} {\bibfnamefont {M.}~\bibnamefont
  {Nowak}},\ }\href@noop {} {\emph {\bibinfo {title} {{Evolutionary
  Dynamics}}}}\ (\bibinfo  {publisher} {Harvard University Press},\ \bibinfo
  {address} {Cambridge, MA},\ \bibinfo {year} {2006})\BibitemShut {NoStop}%
\bibitem [{\citenamefont {{Van Kampen}}(2007)}]{VanKampen2007}%
  \BibitemOpen
  \bibfield  {author} {\bibinfo {author} {\bibfnamefont {N.~G.}\ \bibnamefont
  {{Van Kampen}}},\ }\href
  {papers://1e8b3b69-4952-4922-8c71-6cef5ec0d747/Paper/p45} {\emph {\bibinfo
  {title} {{Stochastic Process in Physics and Chemistry}}}},\ \bibinfo
  {edition} {3rd}\ ed.\ (\bibinfo  {publisher} {Elsevier},\ \bibinfo {address}
  {Amsterdam},\ \bibinfo {year} {2007})\BibitemShut {NoStop}%
\bibitem [{\citenamefont {Laird}\ and\ \citenamefont
  {Schamp}(2009)}]{Laird2009}%
  \BibitemOpen
  \bibfield  {author} {\bibinfo {author} {\bibfnamefont {R.}~\bibnamefont
  {Laird}}\ and\ \bibinfo {author} {\bibfnamefont {B.~S.}\ \bibnamefont
  {Schamp}},\ }\href {\doibase 10.1016/j.jtbi.2008.09.017} {\bibfield
  {journal} {\bibinfo  {journal} {J. Theor. Biol.}\ }\textbf {\bibinfo {volume}
  {256}},\ \bibinfo {pages} {90} (\bibinfo {year} {2009})}\BibitemShut
  {NoStop}%
\bibitem [{\citenamefont {Li}\ \emph {et~al.}(2011)\citenamefont {Li},
  \citenamefont {Dong},\ and\ \citenamefont {Yang}}]{Li2011}%
  \BibitemOpen
  \bibfield  {author} {\bibinfo {author} {\bibfnamefont {Y.}~\bibnamefont
  {Li}}, \bibinfo {author} {\bibfnamefont {L.}~\bibnamefont {Dong}}, \ and\
  \bibinfo {author} {\bibfnamefont {G.}~\bibnamefont {Yang}},\ }\href {\doibase
  10.1016/j.physa.2011.08.019} {\bibfield  {journal} {\bibinfo  {journal}
  {Physica A}\ }\textbf {\bibinfo {volume} {391}},\ \bibinfo {pages} {125}
  (\bibinfo {year} {2011})}\BibitemShut {NoStop}%
\bibitem [{\citenamefont {Klemm}\ and\ \citenamefont
  {Bornholdt}(2005)}]{Klemm2005}%
  \BibitemOpen
  \bibfield  {author} {\bibinfo {author} {\bibfnamefont {K.}~\bibnamefont
  {Klemm}}\ and\ \bibinfo {author} {\bibfnamefont {S.}~\bibnamefont
  {Bornholdt}},\ }\href {\doibase 10.1103/PhysRevE.72.055101} {\bibfield
  {journal} {\bibinfo  {journal} {Phys. Rev. E}\ }\textbf {\bibinfo {volume}
  {72}},\ \bibinfo {pages} {055101} (\bibinfo {year} {2005})}\BibitemShut
  {NoStop}%
\bibitem [{\citenamefont {Case}\ \emph {et~al.}(2010)\citenamefont {Case},
  \citenamefont {Durney}, \citenamefont {Pleimling},\ and\ \citenamefont
  {Zia}}]{Case2010}%
  \BibitemOpen
  \bibfield  {author} {\bibinfo {author} {\bibfnamefont {S.~O.}\ \bibnamefont
  {Case}}, \bibinfo {author} {\bibfnamefont {C.~H.}\ \bibnamefont {Durney}},
  \bibinfo {author} {\bibfnamefont {M.}~\bibnamefont {Pleimling}}, \ and\
  \bibinfo {author} {\bibfnamefont {R.~K.~P.}\ \bibnamefont {Zia}},\ }\href
  {http://stacks.iop.org/0295-5075/92/i=5/a=58003} {\bibfield  {journal}
  {\bibinfo  {journal} {Europhys. Lett.}\ }\textbf {\bibinfo {volume} {92}},\
  \bibinfo {pages} {58003} (\bibinfo {year} {2010})}\BibitemShut {NoStop}%
\bibitem [{\citenamefont {Durney}\ \emph {et~al.}(2011)\citenamefont {Durney},
  \citenamefont {Case}, \citenamefont {Pleimling},\ and\ \citenamefont
  {Zia}}]{Durney2011}%
  \BibitemOpen
  \bibfield  {author} {\bibinfo {author} {\bibfnamefont {C.~H.}\ \bibnamefont
  {Durney}}, \bibinfo {author} {\bibfnamefont {S.~O.}\ \bibnamefont {Case}},
  \bibinfo {author} {\bibfnamefont {M.}~\bibnamefont {Pleimling}}, \ and\
  \bibinfo {author} {\bibfnamefont {R.~K.~P.}\ \bibnamefont {Zia}},\ }\href
  {\doibase 10.1103/PhysRevE.83.051108} {\bibfield  {journal} {\bibinfo
  {journal} {Phys. Rev. E}\ }\textbf {\bibinfo {volume} {83}},\ \bibinfo
  {pages} {051108} (\bibinfo {year} {2011})}\BibitemShut {NoStop}%
\bibitem [{\citenamefont {Durney}\ \emph {et~al.}(2012)\citenamefont {Durney},
  \citenamefont {Case}, \citenamefont {Pleimling},\ and\ \citenamefont
  {Zia}}]{Durney2012}%
  \BibitemOpen
  \bibfield  {author} {\bibinfo {author} {\bibfnamefont {C.~H.}\ \bibnamefont
  {Durney}}, \bibinfo {author} {\bibfnamefont {S.~O.}\ \bibnamefont {Case}},
  \bibinfo {author} {\bibfnamefont {M.}~\bibnamefont {Pleimling}}, \ and\
  \bibinfo {author} {\bibfnamefont {R.~K.~P.}\ \bibnamefont {Zia}},\ }\href
  {\doibase 10.1088/1742-5468/2012/06/P06014} {\bibfield  {journal} {\bibinfo
  {journal} {J. Stat. Phys.}\ }\textbf {\bibinfo {volume} {2012}},\ \bibinfo
  {pages} {P06014} (\bibinfo {year} {2012})}\BibitemShut {NoStop}%
\bibitem [{\citenamefont {Goh}(1977)}]{Goh1977}%
  \BibitemOpen
  \bibfield  {author} {\bibinfo {author} {\bibfnamefont {B.~S.}\ \bibnamefont
  {Goh}},\ }\href {http://www.jstor.org/stable/10.2307/2459985} {\bibfield
  {journal} {\bibinfo  {journal} {Am. Nat.}\ }\textbf {\bibinfo {volume}
  {111}},\ \bibinfo {pages} {135} (\bibinfo {year} {1977})}\BibitemShut
  {NoStop}%
\bibitem [{\citenamefont {Redheffer}(1984)}]{Redheffer1984}%
  \BibitemOpen
  \bibfield  {author} {\bibinfo {author} {\bibfnamefont {R.}~\bibnamefont
  {Redheffer}},\ }\href {http://cat.inist.fr/?aModele=afficheN\&cpsidt=9618974}
  {\bibfield  {journal} {\bibinfo  {journal} {J. Differ. Equations}\ }\textbf
  {\bibinfo {volume} {263}},\ \bibinfo {pages} {245} (\bibinfo {year}
  {1984})}\BibitemShut {NoStop}%
\bibitem [{\citenamefont {Zeeman}(1995)}]{Zeeman1995}%
  \BibitemOpen
  \bibfield  {author} {\bibinfo {author} {\bibfnamefont {M.}~\bibnamefont
  {Zeeman}},\ }\href@noop {} {\bibfield  {journal} {\bibinfo  {journal} {P. Am.
  Math. Soc.}\ }\textbf {\bibinfo {volume} {123}},\ \bibinfo {pages} {87}
  (\bibinfo {year} {1995})}\BibitemShut {NoStop}%
\bibitem [{\citenamefont {Takeuchi}(1996)}]{Takeuchi}%
  \BibitemOpen
  \bibfield  {author} {\bibinfo {author} {\bibfnamefont {Y.}~\bibnamefont
  {Takeuchi}},\ }\href@noop {} {\emph {\bibinfo {title} {{Global Dynamical
  Properties of Lotka-Volterra Systems}}}}\ (\bibinfo  {publisher} {World
  Scientific},\ \bibinfo {year} {1996})\BibitemShut {NoStop}%
\bibitem [{\citenamefont {Chauvet}\ \emph {et~al.}(2002)\citenamefont
  {Chauvet}, \citenamefont {Paullet}, \citenamefont {Previte},\ and\
  \citenamefont {Walls}}]{Chauvet2002}%
  \BibitemOpen
  \bibfield  {author} {\bibinfo {author} {\bibfnamefont {E.}~\bibnamefont
  {Chauvet}}, \bibinfo {author} {\bibfnamefont {J.}~\bibnamefont {Paullet}},
  \bibinfo {author} {\bibfnamefont {J.}~\bibnamefont {Previte}}, \ and\
  \bibinfo {author} {\bibfnamefont {Z.}~\bibnamefont {Walls}},\ }\href
  {http://www.jstor.org/stable/10.2307/3219158} {\bibfield  {journal} {\bibinfo
   {journal} {Math. Magazine}\ }\textbf {\bibinfo {volume} {75}},\ \bibinfo
  {pages} {243} (\bibinfo {year} {2002})}\BibitemShut {NoStop}%
\bibitem [{\citenamefont {Cheon}(2003)}]{Cheon2003}%
  \BibitemOpen
  \bibfield  {author} {\bibinfo {author} {\bibfnamefont {T.}~\bibnamefont
  {Cheon}},\ }\href {\doibase 10.1103/PhysRevLett.90.258105} {\bibfield
  {journal} {\bibinfo  {journal} {Phys. Rev. Lett.}\ }\textbf {\bibinfo
  {volume} {90}},\ \bibinfo {pages} {258105} (\bibinfo {year}
  {2003})}\BibitemShut {NoStop}%
\bibitem [{\citenamefont {{The Big Bang Theory}}(2008)}]{BBT2008}%
  \BibitemOpen
  \bibfield  {author} {\bibinfo {author} {\bibnamefont {{The Big Bang
  Theory}}},\ }\href {http://www.imdb.com/title/tt1256039/} {\enquote {\bibinfo
  {title} {\emph{The Lizard-Spock Expansion}}}\ }\bibinfo {howpublished}
  {Season 2, Episode 8} (\bibinfo {year} {2008})\BibitemShut {NoStop}%
\bibitem [{\citenamefont {Zia}(2010)}]{Zia2011}%
  \BibitemOpen
  \bibfield  {author} {\bibinfo {author} {\bibfnamefont {R.~K.~P.}\
  \bibnamefont {Zia}},\ }\href {http://arxiv.org/abs/1101.0018} {\bibfield
  {journal} {\bibinfo  {journal} {e-print arXiv:1101.0018}\ } (\bibinfo {year}
  {2010})}\BibitemShut {NoStop}%
\bibitem [{\citenamefont {{Meyer}}(2000)}]{Meyer2000}%
  \BibitemOpen
  \bibfield  {author} {\bibinfo {author} {\bibfnamefont {C.~D.}\ \bibnamefont
  {{Meyer}}},\ }\href@noop {} {\emph {\bibinfo {title} {{Matrix Analysis and
  Applied Linear Algebra}}}},\ \bibinfo {edition} {3rd}\ ed.\ (\bibinfo
  {publisher} {SIAM},\ \bibinfo {address} {Philadelphia},\ \bibinfo {year}
  {2000})\BibitemShut {NoStop}%
\bibitem [{\citenamefont {{Cullis}}(1913)}]{Cullis1913}%
  \BibitemOpen
  \bibfield  {author} {\bibinfo {author} {\bibfnamefont {C.~E.}\ \bibnamefont
  {{Cullis}}},\ }\href@noop {} {\emph {\bibinfo {title} {{Matrices and
  Determinoids}}}},\ \bibinfo {edition} {1st}\ ed.,\ Vol.\ \bibinfo {volume} {I
  and II}\ (\bibinfo  {publisher} {Cambridge University Press},\ \bibinfo
  {address} {Cambridge},\ \bibinfo {year} {1913})\BibitemShut {NoStop}%
\bibitem [{\citenamefont {Berr}\ \emph {et~al.}(2009)\citenamefont {Berr},
  \citenamefont {Reichenbach}, \citenamefont {Schottenloher},\ and\
  \citenamefont {Frey}}]{Berr2009}%
  \BibitemOpen
  \bibfield  {author} {\bibinfo {author} {\bibfnamefont {M.}~\bibnamefont
  {Berr}}, \bibinfo {author} {\bibfnamefont {T.}~\bibnamefont {Reichenbach}},
  \bibinfo {author} {\bibfnamefont {M.}~\bibnamefont {Schottenloher}}, \ and\
  \bibinfo {author} {\bibfnamefont {E.}~\bibnamefont {Frey}},\ }\href {\doibase
  10.1103/PhysRevLett.102.048102} {\bibfield  {journal} {\bibinfo  {journal}
  {Phys. Rev. Lett.}\ }\textbf {\bibinfo {volume} {102}},\ \bibinfo {pages}
  {048102} (\bibinfo {year} {2009})}\BibitemShut {NoStop}%
\bibitem [{\citenamefont {Hawick}(2011)}]{Hawick2011}%
  \BibitemOpen
  \bibfield  {author} {\bibinfo {author} {\bibfnamefont {K.}~\bibnamefont
  {Hawick}},\ }\href {http://www.massey.ac.nz/~kahawick/cstn/129/cstn-129.pdf}
  {\emph {\bibinfo {title} {{Cycles, Diversity and Competition in
  Rock-Paper-Scissors-Lizard-Spock Spatial Game Agent Simulations}}}},\
  \bibinfo {type} {Tech. Rep.}\ (\bibinfo  {institution} {Massey University},\
  \bibinfo {year} {2011})\BibitemShut {NoStop}%
\bibitem [{\citenamefont {Krapivsky}\ \emph {et~al.}(2010)\citenamefont
  {Krapivsky}, \citenamefont {Redner},\ and\ \citenamefont
  {Ben-Naim}}]{Krapivsky2010}%
  \BibitemOpen
  \bibfield  {author} {\bibinfo {author} {\bibfnamefont {P.~L.}\ \bibnamefont
  {Krapivsky}}, \bibinfo {author} {\bibfnamefont {S.}~\bibnamefont {Redner}}, \
  and\ \bibinfo {author} {\bibfnamefont {E.}~\bibnamefont {Ben-Naim}},\
  }\href@noop {} {\emph {\bibinfo {title} {{A Kinetic View of Statistical
  Physics}}}},\ \bibinfo {edition} {1st}\ ed.\ (\bibinfo  {publisher}
  {Cambridge University Press},\ \bibinfo {address} {Cambridge},\ \bibinfo
  {year} {2010})\BibitemShut {NoStop}%
\bibitem [{\citenamefont {Frachebourg}\ and\ \citenamefont
  {Krapivsky}(1998)}]{Frachebourg1998}%
  \BibitemOpen
  \bibfield  {author} {\bibinfo {author} {\bibfnamefont {L.}~\bibnamefont
  {Frachebourg}}\ and\ \bibinfo {author} {\bibfnamefont {P.~L.}\ \bibnamefont
  {Krapivsky}},\ }\href {\doibase 10.1088/0305-4470/31/15/001} {\bibfield
  {journal} {\bibinfo  {journal} {J. Phys. A: Math. Gen.}\ }\textbf {\bibinfo
  {volume} {31}},\ \bibinfo {pages} {L287} (\bibinfo {year}
  {1998})}\BibitemShut {NoStop}%
\bibitem [{\citenamefont {Antal}\ and\ \citenamefont
  {Scheuring}(2006)}]{Antal2006}%
  \BibitemOpen
  \bibfield  {author} {\bibinfo {author} {\bibfnamefont {T.}~\bibnamefont
  {Antal}}\ and\ \bibinfo {author} {\bibfnamefont {I.}~\bibnamefont
  {Scheuring}},\ }\href {\doibase 10.1007/s11538-006-9061-4} {\bibfield
  {journal} {\bibinfo  {journal} {Bull. Math. Biol.}\ }\textbf {\bibinfo
  {volume} {68}},\ \bibinfo {pages} {1923} (\bibinfo {year}
  {2006})}\BibitemShut {NoStop}%
\bibitem [{\citenamefont {Parker}\ and\ \citenamefont
  {Kamenev}(2009)}]{Parker2009}%
  \BibitemOpen
  \bibfield  {author} {\bibinfo {author} {\bibfnamefont {M.}~\bibnamefont
  {Parker}}\ and\ \bibinfo {author} {\bibfnamefont {A.}~\bibnamefont
  {Kamenev}},\ }\href {\doibase 10.1103/PhysRevE.80.021129} {\bibfield
  {journal} {\bibinfo  {journal} {Phys. Rev. E}\ }\textbf {\bibinfo {volume}
  {80}},\ \bibinfo {pages} {021129} (\bibinfo {year} {2009})}\BibitemShut
  {NoStop}%
\bibitem [{\citenamefont {Gillespie}(1976)}]{Gillespie1976}%
  \BibitemOpen
  \bibfield  {author} {\bibinfo {author} {\bibfnamefont {D.~T.}\ \bibnamefont
  {Gillespie}},\ }\href@noop {} {\bibfield  {journal} {\bibinfo  {journal} {J.
  Comp. Phys.}\ }\textbf {\bibinfo {volume} {22}},\ \bibinfo {pages} {403}
  (\bibinfo {year} {1976})}\BibitemShut {NoStop}%
\bibitem [{\citenamefont {Cremer}\ \emph {et~al.}(2009)\citenamefont {Cremer},
  \citenamefont {Reichenbach},\ and\ \citenamefont {Frey}}]{Cremer2009}%
  \BibitemOpen
  \bibfield  {author} {\bibinfo {author} {\bibfnamefont {J.}~\bibnamefont
  {Cremer}}, \bibinfo {author} {\bibfnamefont {T.}~\bibnamefont {Reichenbach}},
  \ and\ \bibinfo {author} {\bibfnamefont {E.}~\bibnamefont {Frey}},\ }\href
  {\doibase 10.1088/1367-2630/11/9/093029} {\bibfield  {journal} {\bibinfo
  {journal} {New J. Phys.}\ }\textbf {\bibinfo {volume} {11}},\ \bibinfo
  {pages} {093029} (\bibinfo {year} {2009})}\BibitemShut {NoStop}%
\bibitem [{\citenamefont {Frey}(2010)}]{Frey2010}%
  \BibitemOpen
  \bibfield  {author} {\bibinfo {author} {\bibfnamefont {E.}~\bibnamefont
  {Frey}},\ }\href {\doibase 10.1016/j.physa.2010.02.047} {\bibfield  {journal}
  {\bibinfo  {journal} {Physica A}\ }\textbf {\bibinfo {volume} {389}},\
  \bibinfo {pages} {4265} (\bibinfo {year} {2010})}\BibitemShut {NoStop}%
\bibitem [{Kne()}]{Knebel2013}%
  \BibitemOpen
  \href@noop {} {}\bibinfo {howpublished} {Manuscript in
  preparation}\BibitemShut {NoStop}%
\end{thebibliography}
\end{document}